\documentclass[aps,prb,superscriptaddress,twocolumn,showpacs,preprintnumbers,amsmath,amssymb]{revtex4}

\usepackage[dvips]{graphicx,color}
\usepackage{dcolumn}
\usepackage{bm}
\usepackage{ulem}
\usepackage{soul}
\begin{document}

\title{Extrinsic spin Hall effects measured with lateral spin valve structures}

\author{Y. Niimi}
\email{niimi@issp.u-tokyo.ac.jp}
\affiliation{Institute for Solid State Physics, University of Tokyo, 5-1-5 Kashiwa-no-ha, Kashiwa, Chiba 277-8581, Japan}
\author{H. Suzuki}
\affiliation{Institute for Solid State Physics, University of Tokyo, 5-1-5 Kashiwa-no-ha, Kashiwa, Chiba 277-8581, Japan}
\author{Y. Kawanishi}
\affiliation{Institute for Solid State Physics, University of Tokyo, 5-1-5 Kashiwa-no-ha, Kashiwa, Chiba 277-8581, Japan}
\author{Y. Omori}
\affiliation{Institute for Solid State Physics, University of Tokyo, 5-1-5 Kashiwa-no-ha, Kashiwa, Chiba 277-8581, Japan}
\author{T. Valet}
\affiliation{In Silicio SAS, 730 rue Ren\'{e} Descartes, 13857 Aix en Provence Cedex 3, France}
\author{A. Fert}
\affiliation{Unit\'{e} Mixte de Physique CNRS/Thales, 91767 Palaiseau France associ\'{e}e \`{a} l'Universit\'{e} de Paris-Sud, 91405 Orsay, France}
\author{Y. Otani}
\affiliation{Institute for Solid State Physics, University of Tokyo, 5-1-5 Kashiwa-no-ha, Kashiwa, Chiba 277-8581, Japan}
\affiliation{RIKEN-CEMS, 2-1 Hirosawa, Wako, Saitama 351-0198, Japan}

\date{February 3, 2014}

\begin{abstract}
The spin Hall effect (SHE), induced by spin-orbit interaction 
in nonmagnetic materials, is one of the promising phenomena 
for conversion between charge and spin currents in spintronic devices. 
The spin Hall (SH) angle is the characteristic parameter of this conversion. 
We have performed experiments of the conversion from spin into charge currents 
by the SHE in lateral spin valve structures. 
We present experimental results on the extrinsic SHEs 
induced by doping nonmagnetic metals, Cu or Ag, with impurities 
having a large spin-orbit coupling, Bi or Pb, as well as results 
on the intrinsic SHE of Au. The SH angle induced by Bi 
in Cu or Ag is negative and particularly large 
for Bi in Cu, 10 times larger than 
the intrinsic SH angle in Au. 
We also observed a large SH angle for CuPb but the SHE signal
disappeared in a few days. Such an aging effect could be related to a fast 
mobility of Pb in Cu and has not been observed in CuBi alloys.
\end{abstract}


\maketitle

\section{Introduction}

The spin Hall effect (SHE) and its inverse (ISHE) are key ingredients 
for spintronic devices since they enable conversion of charge currents 
to and from spin currents without using ferromagnets and 
external magnetic fields~\cite{maekawa_review_book}. 
One of the typical examples of utilizing the ISHE is a detection of 
a spin dependent chemical potential arising from the spin Seebeck 
effect~\cite{uchida_nature_2008,uchida_nat_mater_2010,uchida_nat_mater_2011,kirihara_nat_mater_2012,chien_prl_2013,saitoh_prl_2013}. 
The spin Seebeck effect converts heat into spin current, and the generated 
spin current can be electrically detected by the ISHEs of 
Pt~\cite{uchida_nature_2008,uchida_nat_mater_2010,uchida_nat_mater_2011,kirihara_nat_mater_2012} and Au~\cite{chien_prl_2013,saitoh_prl_2013}. 
Magnetization switching with a CoFeB/Ta bilayer film is 
another example of utilizing the SHE~\cite{liu_science_2012}. 
A pure spin current, flow of only spin angular momentum 
without charge current, is generated by the SHE of Ta, 
and induces a spin transfer torque in the ferromagnetic layer. 
To realize the detection of the spin Seebeck effect as well as 
the magnetization switching, the ISHEs and SHEs of simple metals 
such as Pt~\cite{uchida_nature_2008,uchida_nat_mater_2010,uchida_nat_mater_2011,kirihara_nat_mater_2012,liu_prl_2012}, 
Au~\cite{chien_prl_2013,saitoh_prl_2013}, and 
Ta~\cite{liu_science_2012} have been mainly used. 
Among them, Pt has been widely believed to be the best SHE material 
exhibiting a large spin Hall (SH) angle which represents 
the conversion yield between charge and spin currents. 
However it is a costly metal, unsuitable for the practical application. 
In addition, the SHEs of 4$d$ and 5$d$ transition metals originate from 
the intrinsic mechanism based on the degeneracy of $d$ orbits 
by spin-orbit (SO) 
coupling~\cite{guo_prl_2008,tanaka_prb_2008,morota_prb_2011}.
This fact indicates that it is difficult to modulate the SH angle artificially 
once the transition metal is fixed. 

There is another type of SHE, the extrinsic SHE induced by scattering 
on impurities with strong SO interaction. 
There are two mechanisms in the extrinsic SHE, 
i.e., the skew scattering~\cite{skew} and the side jump~\cite{side_jump}. 
Unlike the case of the intrinsic SHE, 
the SH angle can be enhanced by changing the combination of 
host and impurity metals. 
According to recent theoretical calculations based on the skew 
scattering~\cite{gradhand_prl_2010,gradhand_prb_2010,fert_prl_2011}, 
some combinations of noble metals and impurities can give rise to very 
large SH angles, for example in Cu or Ag doped with Bi. 
We have experimentally demonstrated the extrinsic SHEs 
induced by Ir~\cite{niimi_prl_2011} and Bi~\cite{niimi_prl_2012} 
impurities in Cu. 
As for Ir-doped Cu, the magnetization switching has already been realized 
using the SHE of CuIr alloys~\cite{ohno_apl_2013}. 
Since Cu is a typical inexpensive metal, Cu-based alloys are desirable 
for future application of spintronic devices. 
In this paper, we studied the SHEs of CuBi, Au, AgBi, and CuPb 
using the spin absorption method in the lateral spin valve (LSV) structure. 
We have already reported that CuBi alloys show a very large SHE and 
the SH angle amounts to $-0.24$. Here we present an exhaustive report 
including the thickness, magnetic field angle, and temperature dependences 
of the SH angle for CuBi alloys, 
and also other combinations of host and impurity metals which are 
predicted to have large SH angles.

In the following session, we explain our method to measure the SHE. 
We also give detailed explanations on how to obtain 
the spin diffusion length and the SH angle from the experimental data, 
since they are the most important physical quantities in the field 
of spintronics. 
In Sec.~III, we present an entirely different method to evaluate 
the spin diffusion length in detail.
After mentioning how to prepare our samples in Sec.~IV, 
we give our experimental results in Sec.~V and then 
summarize the results in Sec.~VI.

\section{Spin absorption method}

\begin{figure}
\begin{center}
\includegraphics[width=8.5cm]{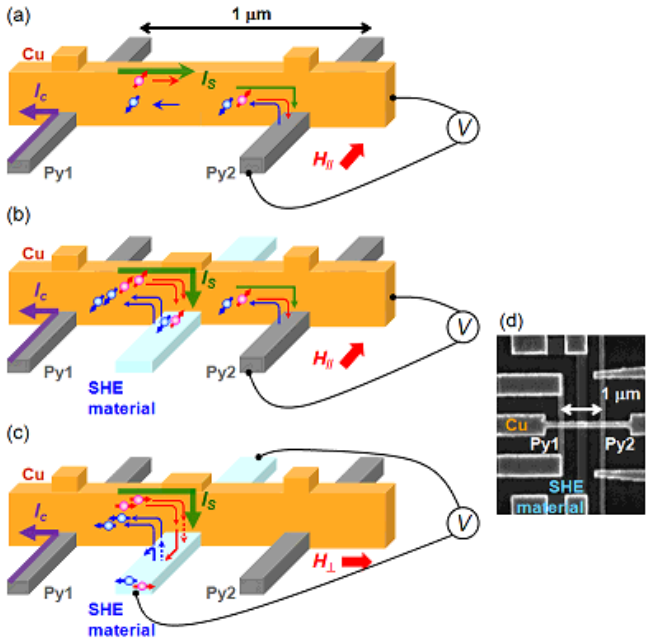}
\caption{(Color online) Spin Hall device in a LSV structure. (a) Schematic of a reference spin valve. The electrochemical potential ($\mu$) distributions of spin-up and spin-down electrons near the interface between Py and Cu are superimposed on the schematic. (b) Schematic of a spin valve with an insertion of a SHE material. Because of a strong SO interaction of the SHE material, a pure spin current ($I_{\rm S}$) is preferentially absorbed into the SHE material. The magnetic field is applied along the easy direction of the Py wires ($H_{\parallel}$) for the nonlocal spin valve (NLSV) measurement. (c) Schematic of the ISHE measurement. The ISHE in the SHE material deflects spin-up and spin-down electrons $|e|$ ($e$ is the charge of the electron) denoted by spheres with arrows to the same side. Other arrows indicate the electron motion direction. The magnetic field is applied along the hard direction of the Py wires ($H_{\perp}$). (d) A typical scanning electron miscroscopy (SEM) image of the SH device. } \label{fig1}
\end{center}
\end{figure}

In the recent field of spintronics, 
there are several methods to measure the SH angle; 
spin pumping in a microwave cavity~\cite{saitoh_apl_2006}, 
spin pumping with coplanar 
waveguides~\cite{hoffmann_prl_2010,hoffmann_prb_2010}, 
spin transfer torque induced ferromagnetic 
resonance~\cite{liu_prl_2011,kondou_apex_2012}, 
SH magnetoresistance~\cite{nakayama_prl_2013}, 
transport measurements with a Hall cross 
structure~\cite{valenzuela_nature_2006,seki_nat_mater_2008,hoffmann_prl_2009,bogu_prl_2010,takanashi_mrs_2012}, 
and spin absorption in a LSV 
structure~\cite{morota_prb_2011,niimi_prl_2011,niimi_prl_2012}. 
In this section, we focus on the spin absorption method 
as shown in Fig.~\ref{fig1}. 
One of the advantages of this method is that 
not only the SH angle but also the spin diffusion length, 
which is a crucial quantity to determine the SH angle, can be determined 
on the same device. 
In addition, the spin absorption method is valid 
for large SO (or short spin diffusion length) 
materials which in general have large SH angles.
To obtain the SH angle and the spin diffusion length, 
we use two different models: 
(i) the one-dimensional (1D) spin diffusion model 
developed by Takahashi and 
Maekawa~\cite{takahashi_prb_2003,takahashi_review_2008}, and 
(ii) the three-dimensional (3D) spin diffusion model 
based on an extension of the Valet-Fert 
formalism~\cite{valet_prb_1993,niimi_prl_2012}. 
The 3D model was originally introduced in Ref.~\onlinecite{niimi_prl_2012} 
to solve a controversial issue about the shunting factor 
used in the 1D model~\cite{morota_prb_2011,niimi_prl_2011,liu_arxiv_2011}. 
As detailed later on, in the 3D model, the shunting factor is 
automatically taken into account when 
the SH angle is evaluated. 
In the following subsections, we explain the two models in detail.

\subsection{1D model}

A LSV consists of two ferromagnetic wires 
and a nonmagnetic wire which bridges the two ferromagnets as shown 
in Fig.~\ref{fig1}(a). 
In the present paper, we use permalloy (Py; Ni$_{81}$Fe$_{19}$) 
as a ferromagnet 
and Cu as a nonmagnetic material (except for Fig.~\ref{fig8}). 
When an electric charge current 
$I_{\rm C} \equiv I_{\uparrow} + I_{\downarrow}$ is injected 
from one of the ferromagnets [Py1 in Fig.~\ref{fig1}(a)] 
into the nonmagnetic material, 
nonequilibrium spin accumulation is generated at the interface and 
is relaxed within a certain length, so-called spin diffusion length. 
Within the spin diffusion length, a pure spin current, 
which is defined as $I_{\rm S} \equiv I_{\uparrow} - I_{\downarrow}$, 
can flow only on the right side of the nonmagnetic wire.
Here $I_{\uparrow}$ and $I_{\downarrow}$ are 
spin-up and spin-down currents, respectively.
The nonequilibrium spin accumulation 
can be detected as a nonlocal voltage $V_{\rm S}$ 
using the other ferromagnetic wire 
[Py2 in Fig.~\ref{fig1}(a)]. 
The detected voltage depends on the magnetization of the 
two ferromagnetic wires, i.e., parallel or antiparallel state. 
The difference in $V_{\rm S}$ between the parallel and antiparallel states, 
i.e., $\Delta V_{\rm S}$ is proportional to the spin accumulation at 
the position of the Py2 detector.
The magnetic field in this case is applied along 
the easy direction of the ferromagnetic wires ($H_{\parallel}$).
As detailed in Ref.~\onlinecite{wakamura_apex_2011}, 
we can determine the spin diffusion lengths 
of Py ($\lambda_{\rm F}$) and Cu ($\lambda_{\rm N}$) 
as well as the spin polarization of Py ($p_{\rm F}$) by 
plotting $\Delta V_{\rm S}$ as a function of the distance ($L$) 
between Py1 and Py2.
In the present study, 
$\lambda_{\rm F}=5$~nm~\cite{bass_jmmm_1997,dubois_prb_1999,niimi_prl_2013}, 
$\lambda_{\rm N}=1.3$~$\mu$m~\cite{wakamura_apex_2011}, 
and $p_{\rm F} = 0.23$~\cite{niimi_prl_2011,niimi_prl_2012} 
at $T=10$~K.

When a SHE material is inserted just in the middle of Py1 and Py2, 
the pure spin current generated from Py1 is partly absorbed into 
the SHE middle wire because of its strong SO interaction, as shown in 
Fig.~\ref{fig1}(b). 
As a result, the spin accumulation detected at Py2 is reduced. 
This reduction, i.e., the spin absorption rate $\eta$, can be 
expressed as follows~\cite{niimi_prl_2011,niimi_prl_2012}:
\begin{widetext}
\begin{align}
\eta \equiv \frac{\Delta R_{\rm S}^{\rm with}}{\Delta R_{\rm S}^{\rm without}} = \frac{2Q_{\rm M} \left\{ \sinh (L/\lambda_{\rm N}) + 2 Q_{\rm F} \exp(L/\lambda_{\rm N})  +2 Q_{\rm F}^{2} \exp(L/\lambda_{\rm N})  \right\} }
{\left\{ \cosh (L/\lambda_{\rm N})-1 \right\} + 2Q_{\rm M} \sinh(L/\lambda_{\rm N}) +2Q_{\rm F}\left\{ \exp(L/\lambda_{\rm N}) (1+Q_{\rm F})(1+2Q_{\rm M}) -1 \right\} } \label{eq2}
\end{align}
\end{widetext}
where $\Delta R_{\rm S}^{\rm with}$ and $\Delta R_{\rm S}^{\rm without}$ are 
the spin accumulation signals ($\Delta V_{\rm S}$ divided 
by the injection current $I_{\rm C}$) with and without 
the SHE middle wire, respectively.
$Q_{\rm F}$ and $Q_{\rm M}$ are defined as 
$R_{\rm F}/R_{\rm N}$, and $R_{\rm M}/R_{\rm N}$, where 
$R_{\rm F}$, $R_{\rm N}$, and $R_{\rm M}$ are the spin resistances of 
Py, Cu, and the middle wire, respectively~\cite{note_spin_resistance}. 
Since only the spin diffusion length $\lambda_{\rm M}$ 
of the SHE middle wire is left as an unknown parameter in Eq.~(\ref{eq2}), 
it can be obtained by measuring $\eta$ experimentally.

In order to measure the SHE with this device, 
we need to apply the magnetic field along the hard direction 
of the Py wires ($H_{\perp}$), as shown in Fig.~\ref{fig1}(c). 
This is related to the fact that 
the charge current $I_{\rm C}$ due to the ISHE is proportional to 
the cross product of $I_{\rm S}$ and the direction of spin. 
In this type of SH device, 
$I_{\rm S}$ is absorbed into the SHE material perpendicularly 
[see Fig.~\ref{fig1}(c)]. 
Thus, to obtain a Hall voltage due to the ISHE ($\Delta V_{\rm ISHE}$), 
the direction of spin has to be aligned 
along the hard direction of the Py wires. 
Based on the 1D spin diffusion model, 
the SH resistivity $\rho_{\rm SHE}$, which is directly related to 
the SH angle, can be written as follows~\cite{takahashi_review_2008,niimi_prl_2011,niimi_prl_2012}: 
\begin{eqnarray}
\rho_{\rm SHE} = \Delta R_{\rm ISHE} 
\frac{w_{\rm M}}{x} \frac{I_{\rm C}}{\bar{I_{\rm S}}}
\label{eq3}
\end{eqnarray}
where $\Delta R_{\rm ISHE} (\equiv \Delta V_{\rm ISHE}/I_{\rm C})$ 
is the amplitude of the ISHE resistance and 
${\bar{I_{\rm S}}}/I_{\rm C}$ is defined as
\begin{widetext}
\begin{align}
\frac{\bar{I_{\rm S}}}{I_{\rm C}} 
= \frac{\lambda_{\rm M}}{t_{\rm M}}\frac{\left( 1-\exp(-t_{\rm M}/\lambda_{\rm M}) \right)^{2}}{1-\exp(-2t_{\rm M}/\lambda_{\rm M})} \frac{2 p_{\rm F} Q_{\rm F} \left\{ \sinh \left( L/2\lambda_{\rm N} \right) + Q_{\rm F}\exp(L/2\lambda_{\rm N}) \right\} } { \left\{ \cosh (L/\lambda_{\rm N})-1 \right\} + 2Q_{\rm M} \sinh(L/\lambda_{\rm N}) +2Q_{\rm F}\left\{ \exp(L/\lambda_{\rm N}) (1+Q_{\rm F})(1+2Q_{\rm M}) -1 \right\} }. \nonumber
\end{align}
\end{widetext}
$\bar{I}_{\rm S}$ and $t_{\rm M}$ are the effective spin current injected 
vertically into the SHE middle wire and the thickness of the middle wire, 
respectively. 
The perpendicularly absorbed pure 
spin current decreases in the SHE material, 
exponentially when $\lambda_{\rm M} < t_{\rm M}$, 
linearly down to zero at the bottom of the SHE wire 
when $\lambda_{\rm M} > t_{\rm M}$. 

The coefficient $x$ in Eq.~(\ref{eq3}) is so-called the shunting factor. 
This factor expresses the shunting by the Cu contact above the SHE material 
and its value, $x \approx 0.36$, can be found by additional measurements 
that have been described in Ref.~\onlinecite{niimi_prl_2011}. 
However, as detailed in Ref.~\onlinecite{liu_arxiv_2011}, 
there was a debate on how to evaluate the shunting factor $x$. 
The evaluation of $x$ is very crucial to determine the SH angle correctly. 
As we will see in the next subsection, 
the shunting is automatically taken into account 
in the 3D finite element analysis.

\subsection{3D model}

The detailed explanation of our 3D model extending 
the 1D Valet-Fert model of spin transport has been presented in the 
Supplemental Material of our previous work~\cite{niimi_prl_2012}. 
Here we focus on how to obtain the spin diffusion length and the SH angle 
with the 3D model.

Numerical calculations based on the 3D version of the Valet-Fert model 
have been performed using SpinFlow~3D. It implements a finite element method 
to solve a discrete formulation of the bulk transport equations, 
supplemented with the interface and boundary conditions. 
In SpinFlow~3D, the interface resistance $r_{b}^{*}$ and 
the spin mixing conductance~\cite{brataas_prl_2000} $g_{\uparrow\downarrow}$ 
between Cu and Py are important parameters. 
In the present case, we take their values from appropriate references; 
$r_{b}^{*}=0.5$~f$\Omega$m$^{2}$ from Ref.~\onlinecite{bass_ass_2009} and 
$g_{\uparrow\downarrow} = 1\times10^{15}$~$\Omega^{-1}$m$^{-2}$ 
from Refs.~\onlinecite{taniguchi_apex_2008} and \onlinecite{ghosh_prl_2012}. 
The interface resistance between Cu and a very weakly doped Cu 
should be very small and we have taken the smallest value found in the
literature, 0.1~f$\Omega$m$^{2}$ [see Ref.~\onlinecite{bass_review_2007}]. 

We first determine the spin polarization $\beta$ 
and the interfacial resistance asymmetry coefficient $\gamma$ values 
in the Valet-Fert model~\cite{valet_prb_1993}
by fitting the nonlocal spin valve (NLSV) signal without any middle wire 
as a function of $L$ with SpinFlow~3D, as we have done with the 1D model. 
In our case, $\beta = \gamma = 0.31$ at 10~K. 
$\beta$ is slightly different from $p_{\rm F}=0.23$ from the 1D model. 
When there is a middle wire in between the two ferromagnetic wires, 
the spin current is partially absorbed into it, leading to 
the reduction of the spin accumulation signal at the detector. 
By choosing an appropriate $\lambda_{\rm M}$ value in SpinFlow~3D, 
we can reproduce $\Delta R_{\rm S}$. 
In a similar way, we can determine the SH angle $\alpha_{\rm H}$. 
We rotate the magnetization direction of the Py wire in SpinFlow~3D 
and put an appropriate $\alpha_{\rm H}$ value for the middle wire. 
As a result, we can reproduce a $R_{\rm ISHE}$ vs $H$ curve in the simulation. 
Here we note that the shunting by the Cu contact is automatically 
taken into account in this 3D finite element calculation.

\section{Weak antilocalization}

\begin{figure}
\begin{center}
\includegraphics[width=7.5cm]{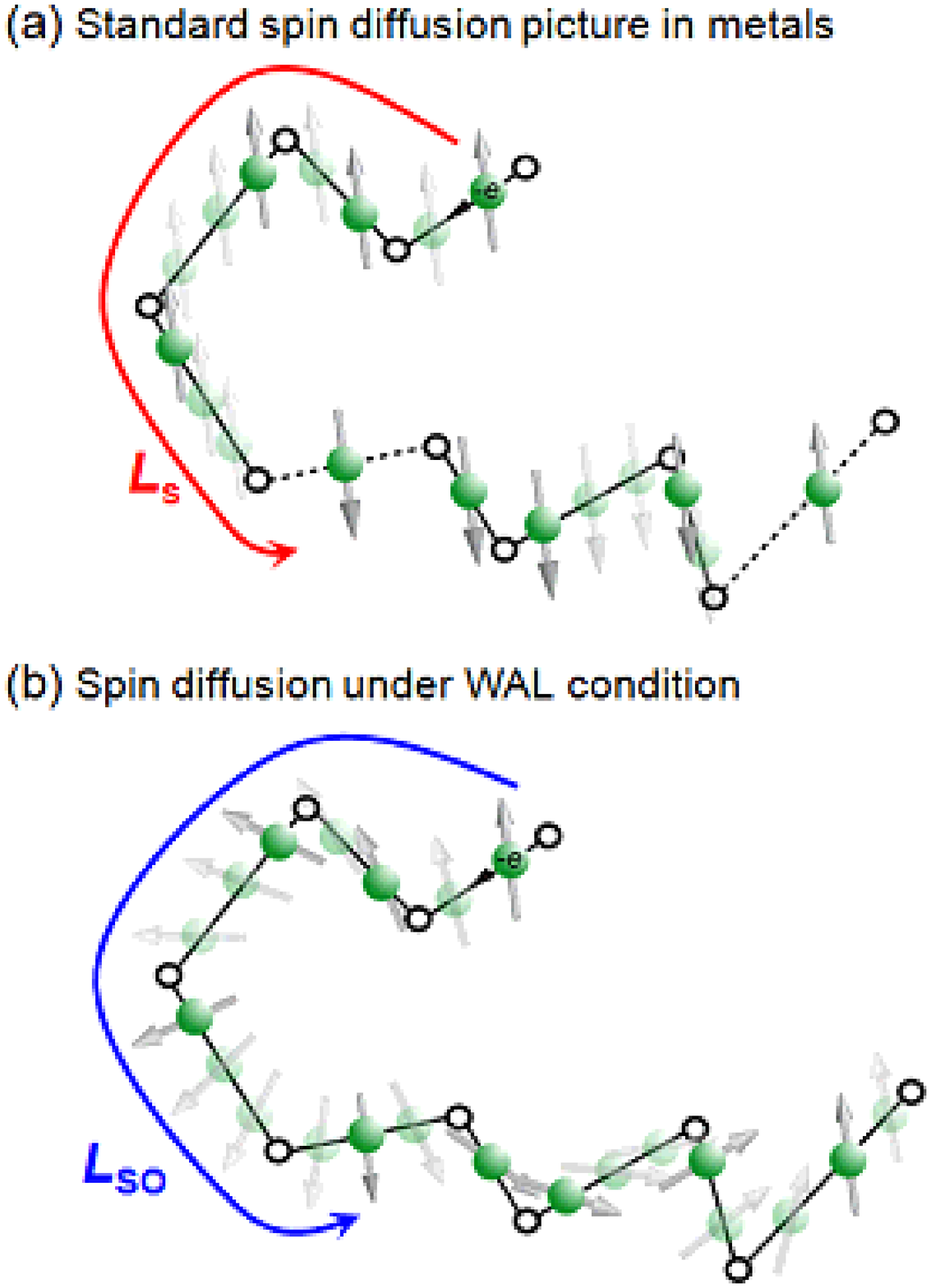}
\caption{(Color online) Schematics of (a) standard spin diffusion picture based on the Elliott-Yafet mechanism~\cite{ey} and (b) spin diffusion under WAL picture. $L_{\rm s}$ and $L_{\rm SO}$ are the spin diffusion length and the SO length, respectively.} \label{fig2}
\end{center}
\end{figure}

As discussed in Sec.~II, the spin absorption method is one of 
the ways to evaluate the spin diffusion length and the SH angle 
on the same device. 
Especially, the evaluation of the spin diffusion length is 
a key issue to estimate the SH angle correctly. 
However, there was a heavy debate about the evaluation of 
the spin diffusion length and the SH angle using this 
method~\cite{liu_arxiv_2011,niimi_prl_2013}. 
In addition, the spin diffusion length of Pt determined with 
the spin absorption method~\cite{morota_prb_2011} 
is always several times larger than 
that obtained with nonmagnet/ferromagnet bilayer 
films~\cite{liu_prl_2011,liu_arxiv_2011}. 
To judge whether the spin diffusion length obtained with 
the spin absorption method is too large or not, 
one needs another approach. 

Weak antilocalization (WAL) is one of the simple ways 
to obtain the spin diffusion length, as already reported 
in previous papers~\cite{bass_review_2007,niimi_prl_2013,bass_review_2013}. 
Weak localization occurs in metallic systems and has been 
used to study decoherence of electrons~\cite{montambaux_book,pierre_prb_2003,niimi_prl_2009,niimi_prb_2010}. 
The principle of this technique relies on constructive interference
of closed electron trajectories which are traveled in opposite
direction (time reversed paths). This leads to an enhancement of the
resistance. The magnetic field $B$ perpendicular to the plane 
destroys these constructive interferences, 
leading to a negative magnetoresistance $R(B)$ 
whose amplitude and width are
directly related to the phase coherence length. 
If there is a non-negligible SO interaction, 
a positive magnetoresistance can be obtained, which 
is referred to as WAL~\cite{hikami_ptp_1980}. 

The dimension of the system is determined with respect to 
the phase coherence length $L_{\varphi}$ and 
the elastic mean free path $l_{e}$. 
Since we deal with nanometer-scale metallic systems, 
$l_{e}$ is in general smaller than all the sample dimensions. 
On the other hand, the inelastic scattering length $L_{\varphi}$ 
can be relatively long for a clean metallic system. 
When $L_{\varphi}$ is larger than the width $w$ and 
the thickness $t$ of the sample but smaller than the length $\ell$,
we call the system ``quasi-1D". 

The WAL peak of quasi-1D wire 
can be fitted by the Hikami-Larkin-Nagaoka 
formula~\cite{montambaux_book,hikami_ptp_1980}:
\begin{widetext}
\begin{eqnarray}
\frac{\Delta R}{R_{\infty}} = \frac{1}{\pi \ell}\frac{R_{\infty}}{\hbar/e^{2}}
\left( \frac{\frac{3}{2}}{\sqrt{\frac{1}{L_{\varphi}^{2}}+\frac{4}{3}\frac{1}{L_{\rm SO}^{2}}+\frac{1}{3}\frac{w^{2}}{l_{B}^{4}}}} 
- \frac{\frac{1}{2}}{\sqrt{\frac{1}{L_{\varphi}^{2}} + \frac{1}{3}\frac{w^{2}}{l_{B}^{4}}}} \right) \label{eq4}
\end{eqnarray}
\end{widetext}
where $\Delta R$, $R_{\infty}$ and $L_{\rm SO}$ are 
the WAL correction factor, the resistance of the wire at high enough field, 
and the SO length, respectively.
$\hbar$ and $l_{B} \equiv \sqrt{\hbar/eB}$ are 
the reduced Plank constant and the magnetic length, respectively. 
In Eq.~(\ref{eq4}), we have only two unknown parameters; 
$L_{\varphi}$ and $L_{\rm SO}$. 
According to the Fermi liquid 
theory~\cite{niimi_prl_2009,niimi_prb_2010,aak_1982}, 
$L_{\varphi}$ depends on temperature ($\propto T^{-1/3}$), 
while $L_{\rm SO}$ is almost constant 
at low temperatures~\cite{pierre_prb_2003}. 

The relation between the SO length and 
the spin diffusion length 
has been theoretically discussed in Ref.~\onlinecite{zutic_review_2004} 
and experimentally verified recently 
by some of the present authors~\cite{niimi_prl_2013}.
The schematics of the two length scales are depicted in Fig.~\ref{fig2}. 
In metallic systems where the Elliott-Yafet mechanism is dominant~\cite{ey},
the following relation can be lead; 
\begin{eqnarray} L_{\rm s} = \frac{\sqrt{3}}{2}L_{\rm SO}. 
\label{eq5} 
\end{eqnarray} 
Since $L_{\rm s}$ is basically equivalent to $\lambda_{\rm N}$ or 
$\lambda_{\rm M}$, we use hereafter 
only $\lambda_{\rm N}$ or $\lambda_{\rm M}$ 
as the spin diffusion length of nonmagnetic metal.

\section{Sample fabrication and experimental setup}

Our SH device is 
based on a LSV structure where a SHE material 
is inserted in between two Py wires 
and bridged by a Cu wire, as shown in Fig.~1. 
Samples were patterned using electron beam lithography onto a thermally 
oxidized silicon substrate coated with polymethyl-methacrylate (PMMA) resist 
for depositions of Py, Cu, Ag, Au, and CuPb alloys, 
or coated with ZEP 520A resist 
for depositions of CuBi and AgBi. 

A pair of Py wires was first deposited 
using an electron beam evaporator under a base pressure of 10$^{-9}$ Torr. 
The width and thickness of the Py wires are 100 and 30~nm, respectively.
The CuBi, AgBi and CuPb middle wires were next deposited 
by magnetron sputtering with Bi-doped Cu and Ag targets 
and Pb-doped Cu targets, respectively. 
The Bi concentrations used in this work were 0\%, 0.3\%, and 0.5\% 
for CuBi, and 0\%, 1\%, and 3\% for AgBi. 
As for CuPb, we used only 0.5\% of Pb in Cu.
We also prepared Au middle wires since the spin diffusion length of Au
is expected to be as long as that of CuBi. 
The Au wires were deposited 
by a Joule heating evaporator using a 99.997\% purity source. 
The width and thickness of CuBi, AgBi and CuPb are 250 and 20~nm 
(except for Fig.~\ref{fig6}) while those of Au are 200 and 20~nm.

The post-baking temperature for the PMMA resist was 
kept below 90~$^{\circ}$C after the deposition of CuBi, AgBi or CuPb alloys. 
Bismuth and lead have low melting temperatures 
(270~$^{\circ}$C and 330~$^{\circ}$C), which oblige us 
to use a much lower post-baking temperature. 
We have confirmed that the post-baking temperature of 90~$^{\circ}$C 
does not change the resistivities of CuBi, AgBi, and CuPb wires. 
Before deposition of a Cu bridge, we performed 
a careful Ar ion beam etching for 30 seconds 
in order to clean the surfaces of Py and the SHE middle wires. 
After the Ar ion etching, the device was moved to another chamber 
without breaking a vacuum and subsequently the Cu bridge was deposited 
by a Joule heating evaporator using a 99.9999\% purity source. 
For comparison, we also prepared similar SH devices but 
bridged by a Ag wire from a 99.999\% purity source. 
Both the width and thickness of Cu (or Ag in Fig.~\ref{fig8}) are 100~nm.

For the WAL samples, we prepared $\sim$ 1~mm
long and 100~nm wide Au wires, and 120~nm wide Cu$_{99.7}$Bi$_{0.3}$ and 
Cu$_{99.5}$Bi$_{0.5}$ wires. The thickness is 20 nm, which is the same as 
in the SHE device.

The measurements have been carried out using an ac lock-in amplifier 
(modulation frequency $f=173$~Hz) and a $^{4}$He flow cryostat. 
In order to obtain a very small WAL signal 
compared to the background resistance, 
we used a bridge circuit as detailed in Ref.~\onlinecite{niimi_prb_2010}.
To check the reproducibility and to evaluate the error bar 
(see Table~\ref{table1}), 
we have measured at least a few different samples from the same batch.

\section{Experimental results and discussions}

\subsection{SHEs of CuBi and Au}

\begin{figure}
\begin{center}
\includegraphics[width=7.5cm]{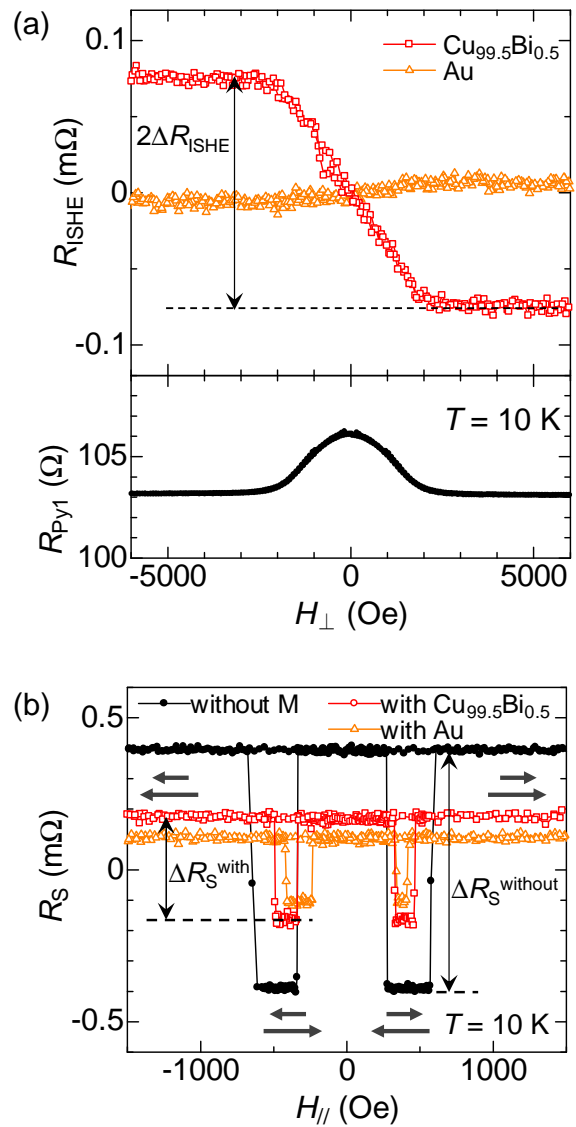}
\caption{(Color online) (a) ISHE resistances ($R_{\rm ISHE}$) of Cu$_{99.5}$Bi$_{0.5}$ and Au measured at $T = 10$~K. The lower panel shows the AMR of Py1, indicating the saturation of the magnetization above $H_{\perp}\sim 2000$~Oe along the hard direction of Py1. (b) NLSV signals ($R_{\rm S}$) measured at $T = 10$~K with a Cu$_{99.5}$Bi$_{0.5}$ wire (open square), with a Au wire (open triangle), and without any middle wire (closed circle). In this case, the magnetic field is aligned along the easy axis of the Py wires ($H_{\parallel}$). The arrows represent the magnetization directions of Py1 (upper arrow) and Py2 (lower arrow). } \label{fig3}
\end{center}
\end{figure}

We first compare two ISHE resistances, i.e., 
$R_{\rm ISHE}$ of Cu$_{99.5}$Bi$_{0.5}$ and Au measured at $T=10$~K 
in Fig.~\ref{fig3}(a). 
As already explained in Sec.~II~A, 
in our device structure,
$R_{\rm ISHE}$ linearly increases with increasing the magnetic field 
and it is saturated above 2000~Oe which is the saturation field of 
the magnetization. 
This saturation field can be confirmed from the anisotropic magnetoresistance 
(AMR) curve of Py1 in the lower panel of Fig.~\ref{fig3}(a).
Although the sign of $R_{\rm ISHE}$ of Cu$_{99.5}$Bi$_{0.5}$ is 
opposite to that of Au, its amplitude 
is more than 10 times larger compared to that of Au.

To evaluate the spin absorption rate $\eta$ and the spin diffusion length of 
the middle wire, we performed NLSV measurements 
with and without the middle wire. 
Figure~\ref{fig3}(b) shows NLSV signals $R_{\rm S}$ with the 
Cu$_{99.5}$Bi$_{0.5}$ and Au middle wires, 
and also $R_{\rm S}$ without any middle wire as a reference signal. 
Apparently, the insertion of the SHE materials in the LSV structure 
induces the reduction in $R_{\rm S}$ detected at Py2. 
By using the 1D and 3D spin transport models, 
the spin diffusion lengths of Cu$_{99.5}$Bi$_{0.5}$ and Au can 
be obtained and are listed in Table~\ref{table1}. 
Both $\lambda_{\rm M}^{\rm 1D}$ and $\lambda_{\rm M}^{\rm 3D}$ 
of Cu$_{99.5}$Bi$_{0.5}$ are almost the same as 
$\lambda_{\rm M}^{\rm 1D}$ and $\lambda_{\rm M}^{\rm 3D}$ of Au, respectively.
As already pointed out in Ref.~\onlinecite{niimi_prl_2012} 
and will be detailed later on, 
when $\lambda_{\rm M} > t_{\rm M}$, the 1D model underestimates 
not only $\lambda_{\rm M}$ but also $\alpha_{\rm H}$ compared to the 3D model. 

\begin{table*}
\caption{Characteristics of various SHE materials measured below 10~K. As for Cu$_{99}$Ir$_{1}$, Pt, and Ta, the raw data were already shown in Ref.~\onlinecite{niimi_prl_2011}, Ref.~\onlinecite{niimi_prl_2013}, and Ref.~\onlinecite{morota_prb_2011}, respectively.}
\label{table1}
\begin{ruledtabular}
\begin{tabular}{cccccccc}
SHE material & method & $\rho$ or $\rho_{\rm imp}$ & $\alpha_{\rm H}^{\rm 3D}$ & $\alpha_{\rm H}^{\rm 1D}$ & $\lambda_{\rm M}^{\rm 3D}$ & $\lambda_{\rm M}^{\rm 1D}$ & $\left( \sqrt{3}/2 \right)L_{\rm SO}$ \\ 
(20~nm) &  & ($\mu\Omega\cdot$cm) &  &  & (nm) & (nm) & (nm) \\
\hline
Au & LSV \& WAL & 4.0 & $0.014(\pm 0.004)$ & $0.010(\pm 0.002)$ & $40(\pm 16)$ & $33(\pm 9)$ & $38(\pm 4)$ \\
Cu$_{99.7}$Bi$_{0.3}$ & LSV \& WAL & 3.2 & $-0.26(\pm 0.11)$ & $-0.11(\pm 0.04)$ & $86(\pm 17)$ & $53(\pm 8)$ & $66(\pm 4)$ \\
Cu$_{99.5}$Bi$_{0.5}$ & LSV \& WAL & 5.1 & $-0.24(\pm 0.09)$ & $-0.12(\pm 0.04)$ & $45(\pm 14)$ & $32(\pm 9)$ & $37(\pm 3)$ \\
Cu$_{99.5}$Pb$_{0.5}$ & LSV & 5.4 & $-0.13(\pm 0.03)$ & $-0.07(\pm 0.02)$ & $53(\pm 15)$ & $36(\pm 7)$ & - \\
Ag$_{99}$Bi$_{1}$ & LSV & 6.8 & $-0.023(\pm 0.006)$ & $-0.016(\pm 0.005)$ & $29(\pm 6)$ & $23(\pm 5)$ & - \\
Cu$_{99}$Ir$_{1}$ [\onlinecite{niimi_prl_2011}] & LSV & 3.1 & $0.023(\pm 0.006)$ & $0.021(\pm 0.06)$ & $36(\pm 7)$ & $27(\pm 5)$ & - \\
Pt [\onlinecite{niimi_prl_2013}] & LSV \& WAL & 10 & $0.024(\pm 0.006)$ & $0.021(\pm 0.005)$ & $10(\pm 2)$ & $11(\pm 2)$ & $10(\pm 2)$ \\ 
Ta [\onlinecite{morota_prb_2011}] & LSV & 330 & $-0.008(\pm 0.002)$ & $-0.004(\pm 0.001)$ & $3(\pm 0.4)$ & $3(\pm 0.4)$ & - \\
\end{tabular}
\end{ruledtabular}
\end{table*}

Concerning the SH angles, 
we obtain $\alpha_{\rm H}^{\rm 3D}=-0.24(\pm 0.09)$ for Cu$_{99.5}$Bi$_{0.5}$, 
if we divide $\rho_{\rm SHE}$ by 
the Bi-induced resistivity 
$\rho_{\rm imp}(\equiv \rho_{\rm CuBi}-\rho_{\rm Cu})$. 
This is based on the fact that the ISHE cannot be detected 
for pure Cu wire and 
the resistivity of pure Cu wire $\rho_{\rm Cu}$ is not negligibly small 
compared to the total resistivity $\rho_{\rm CuBi}$. 
If $\rho_{\rm SHE}$ is divided by $\rho_{\rm CuBi}$, 
the SH angle of Cu$_{99.5}$Bi$_{0.5}$ becomes $-0.11$ as already 
pointed out in Ref.~\onlinecite{niimi_prl_2012}.
On the other hand, $\alpha_{\rm H}^{\rm 3D}$ of 20~nm thick Au is 
$0.014(\pm 0.004)$. This SH angle is consistent with the values 
reported in some previous works with comparable Au 
thicknesses~\cite{hoffmann_prl_2009,hoffmann_prl_2010,hoffmann_prb_2010,bogu_prl_2010,takanashi_mrs_2012}. 

\begin{figure}
\begin{center}
\includegraphics[width=7.5cm]{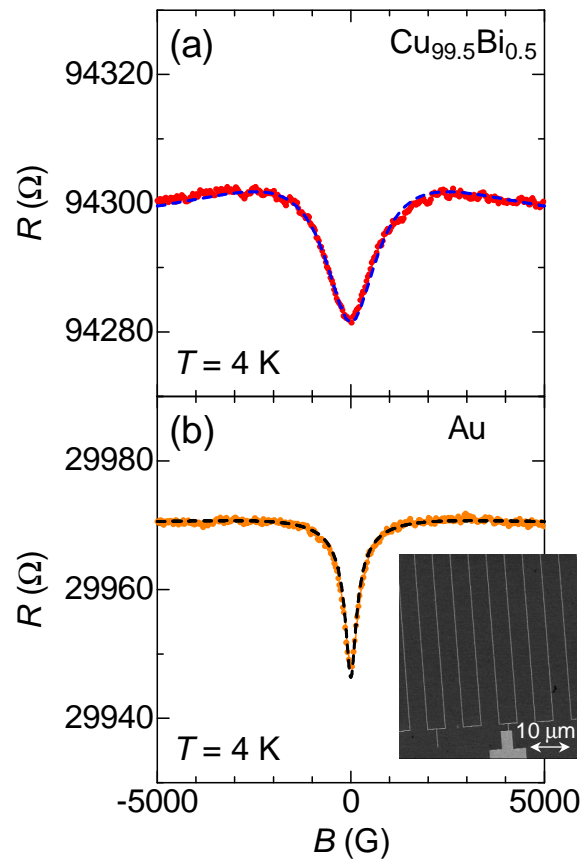}
\caption{(Color online) WAL curves of (a) Cu$_{99.5}$Bi$_{0.5}$ wire ($w=120$~nm, $t=20$~nm, $\ell=1.4$~mm) and (b) Au wire ($w=100$~nm, $t=20$~nm, $\ell=1.2$~mm) measured at $T = 4$~K. The broken lines are the best fits to Eq.~(\ref{eq4}). The inset shows an SEM image of the Au wire. The magnetic field $B$ is applied perpendicular to the plane.} \label{fig4}
\end{center}
\end{figure}


In order to double-check $\lambda_{\rm M}$ obtained from 
the spin absorption measurements, we prepared simple 
Cu$_{99.5}$Bi$_{0.5}$ and Au wires, and performed WAL measurements at $T=4$~K 
as shown in Fig.~\ref{fig4}. 
For both wires, clear positive magnetoresistance is observed, 
which is typical of WAL. 
By fitting the WAL curves with Eq.~(\ref{eq4}), $L_{\rm SO}$ 
can be obtained and converted into 
$\lambda_{\rm M}$ using Eq.~(\ref{eq5}). 
As can be seen in Table~\ref{table1}, 
the obtained $(\sqrt{3}/2)L_{\rm SO}$ of Cu$_{99.7}$Bi$_{0.3}$, 
Cu$_{99.5}$Bi$_{0.5}$ and Au 
are quantitatively consistent with $\lambda_{\rm M}^{\rm 3D}$ from 
the spin absorption measurements. 
Thus, it turns out that the WAL method is valid to evaluate 
the spin diffusion length quantitatively not only for pure 
metals such as Pt~\cite{niimi_prl_2013} and Au but also 
for dilute alloys.

\subsection{Thickness dependence of SHE of CuBi}

\begin{figure}
\begin{center}
\includegraphics[width=7cm]{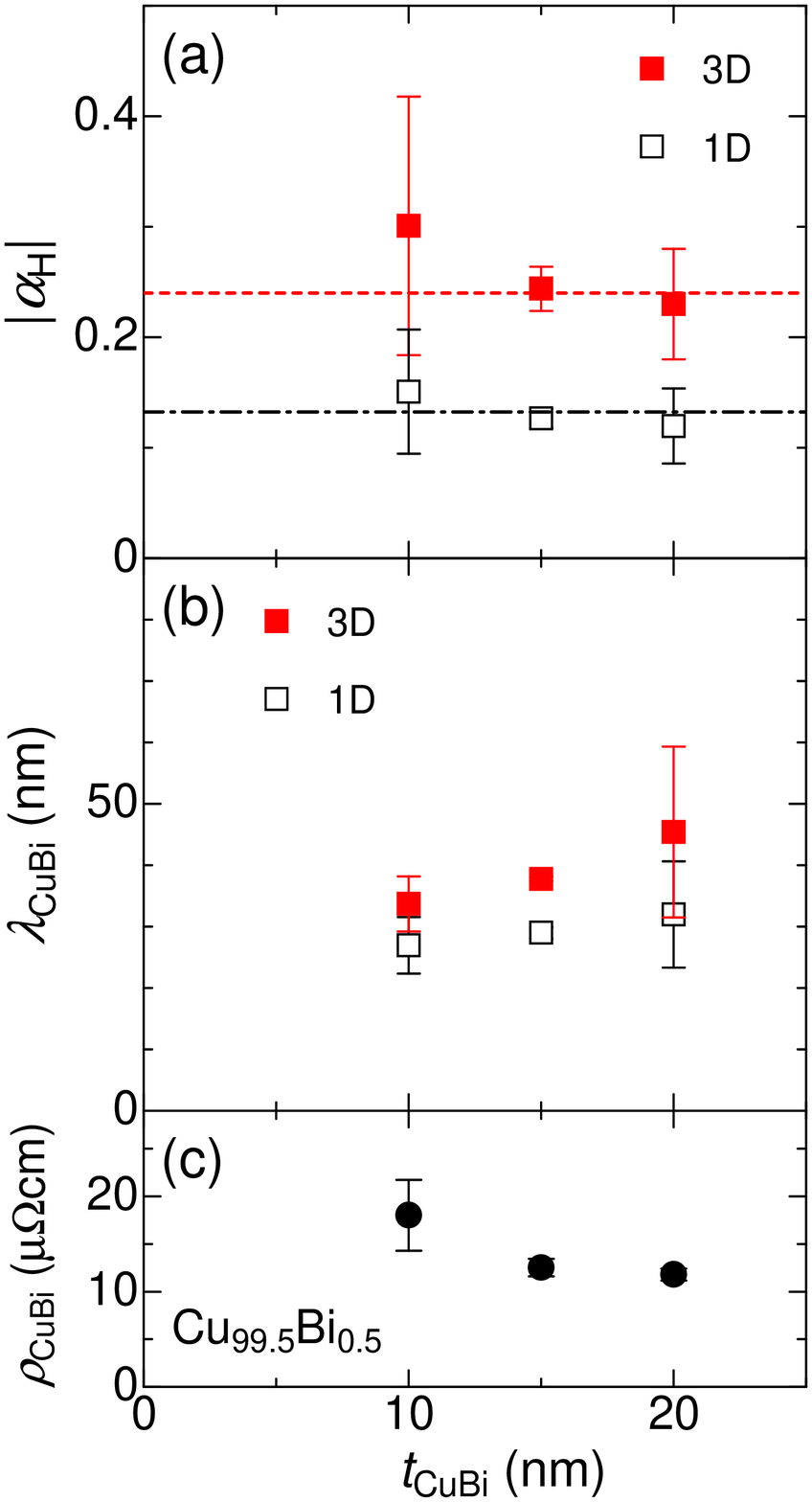}
\caption{(Color online) Thickness dependence of (a) the SH angle, (b) the spin diffusion length, and (c) the resistivity of Cu$_{99.5}$Bi$_{0.5}$ measured at $T=10$~K. The broken and dashed-dotted lines in (a) correspond to $|\alpha_{\rm H}^{\rm 3D}|$ and $|\alpha_{\rm H}^{\rm 1D}|$, respectively.} \label{fig6}
\end{center}
\end{figure}

\begin{figure*}
\begin{center}
\includegraphics[width=17cm]{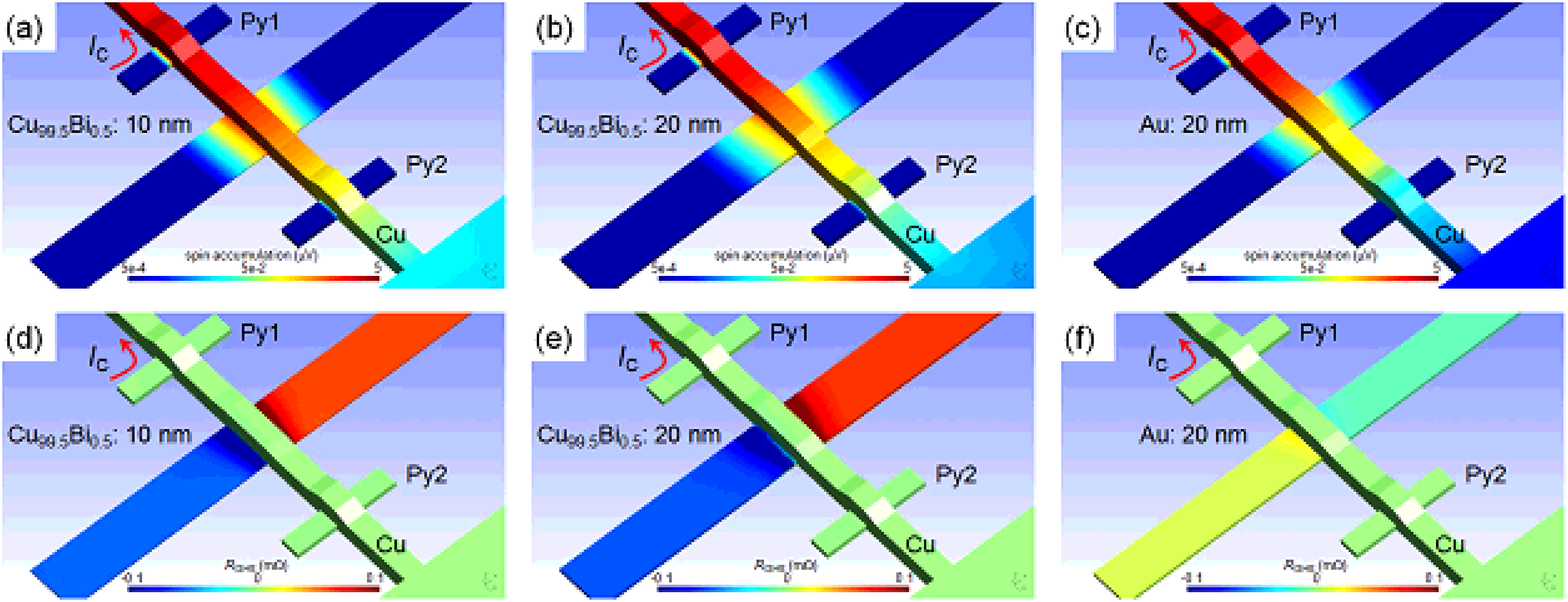}
\caption{(Color online) (a)-(c) 3D mappings of the spin accumulation voltage calculated with SpinFlow~3D for (a) 10~nm thick Cu$_{99.5}$Bi$_{0.5}$, (b) 20~nm thick Cu$_{99.5}$Bi$_{0.5}$, and (c) 20~nm thick Au SH devices. Note that the color scale is a logarithmic scale. (d)-(f) 3D mappings of $R_{\rm ISHE}$ calculated with SpinFlow~3D for (d) 10~nm thick Cu$_{99.5}$Bi$_{0.5}$, (e) 20~nm thick Cu$_{99.5}$Bi$_{0.5}$, and (f) 20~nm thick Au SH devices. The color scale here is a linear scale.} \label{fig7}
\end{center}
\end{figure*}

Next we study the thickness dependence of 
the SHE of Cu$_{99.5}$Bi$_{0.5}$. 
Figure~\ref{fig6} shows the thickness dependence of (a) the SH angle, 
(b) the spin diffusion length, and (c) the resistivity of 
the Cu$_{99.5}$Bi$_{0.5}$ middle wire. 
With decreasing $t_{\rm M}$, the SH angle slightly increases but not 
so drastically compared to the case of 
$\beta$-tungsten~\cite{ralph_apl_2013}. 
If we assume that the SH angle is independent of $t_{\rm M}$ below 20~nm, 
we obtain $\alpha_{\rm H} = -0.24$, which is consistent with 
the value estimated from the 
Bi concentration dependence~\cite{niimi_prl_2012}. 
This almost independent $\alpha_{\rm H}$ with respect to $t_{\rm M}$ 
indicates that the ISHE signal originates 
from the skew scattering from homogenously distributed 
Bi impurities in the Cu wire. 
On the other hand, the spin diffusion length decreases 
with decreasing the thickness as shown in Fig.~\ref{fig6}(b). 
This tendency can be qualitatively understood by 
the resistivity change [see Fig.~\ref{fig6}(c)], 
as reported in our previous work~\cite{niimi_prl_2013} 
where the spin diffusion length of Cu 
is inversely proportional to its resistivity. 

As already shown in Ref.~\onlinecite{niimi_prl_2012}, 
the spreading of the spin accumulation at the side edges 
of the SHE material wire leads to the underestimations 
of $\lambda_{\rm M}$ and $\alpha_{\rm H}$ 
when $\lambda_{\rm M} > t_{\rm M}$. 
In Fig.~\ref{fig7}, we show 3D mappings of 
the spin accumulation voltage for (a) 
10~nm thick Cu$_{99.5}$Bi$_{0.5}$, (b) 20~nm thick Cu$_{99.5}$Bi$_{0.5}$, 
and (c) 20~nm thick Au devices. 
The corresponding $R_{\rm ISHE}$ are also plotted 
in Figs.~\ref{fig7}(d)-\ref{fig7}(f). 
For both Cu$_{99.5}$Bi$_{0.5}$ and Au middle wires, 
the spreading of the spin accumulation can be seen since 
the spin diffusion lengths of Cu$_{99.5}$Bi$_{0.5}$ and Au 
are larger than $t_{\rm M}$. 
This is the reason for the difference between $\lambda_{\rm M}^{\rm 1D}$ and 
$\lambda_{\rm M}^{\rm 3D}$ as well as the difference between 
$\alpha_{\rm H}^{\rm 1D}$ and $\alpha_{\rm H}^{\rm 3D}$. 
Compared to the 20~nm thick Cu$_{99.5}$Bi$_{0.5}$ device, 
the spreading of the spin accumulation is smaller for 
the 10~nm thick Cu$_{99.5}$Bi$_{0.5}$ device 
[see Figs.~\ref{fig7}(a) and \ref{fig7}(b)], 
although $R_{\rm ISHE}$ are almost the same for the two devices 
[see Figs.~\ref{fig7}(d) and \ref{fig7}(e)]. 
This comes from the thickness 
dependence of the spin diffusion length, 
i.e., smaller $\lambda_{\rm M}$ for thinner $t_{\rm M}$, 
as shown in Fig.~\ref{fig6}(b). 

\subsection{SHEs of several different materials}

\begin{figure}
\begin{center}
\includegraphics[width=7.5cm]{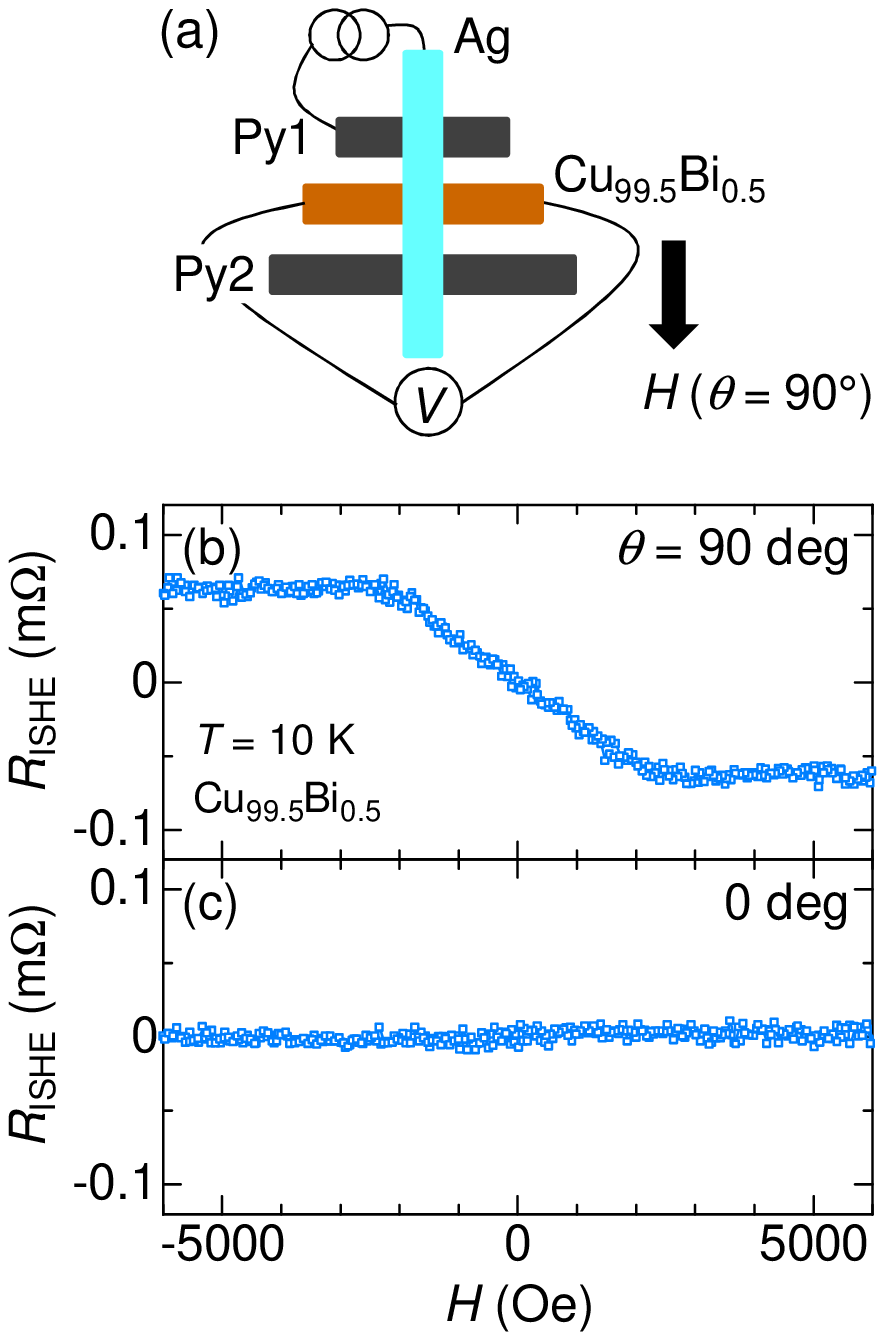}
\caption{(Color online) Substitution of Ag bridge for Cu bridge. (a) Schematic of ISHE measurement. The arrow indicates the applied filed direction. We define $H_{\perp}$ as $\theta=90^{\circ}$. The angle dependence of $R_{\rm ISHE}$ measured at (b) $\theta = 90^{\circ}$, (c) $\theta = 0^{\circ}$. 
} \label{fig8}
\end{center}
\end{figure}

\begin{figure}
\begin{center}
\includegraphics[width=7.5cm]{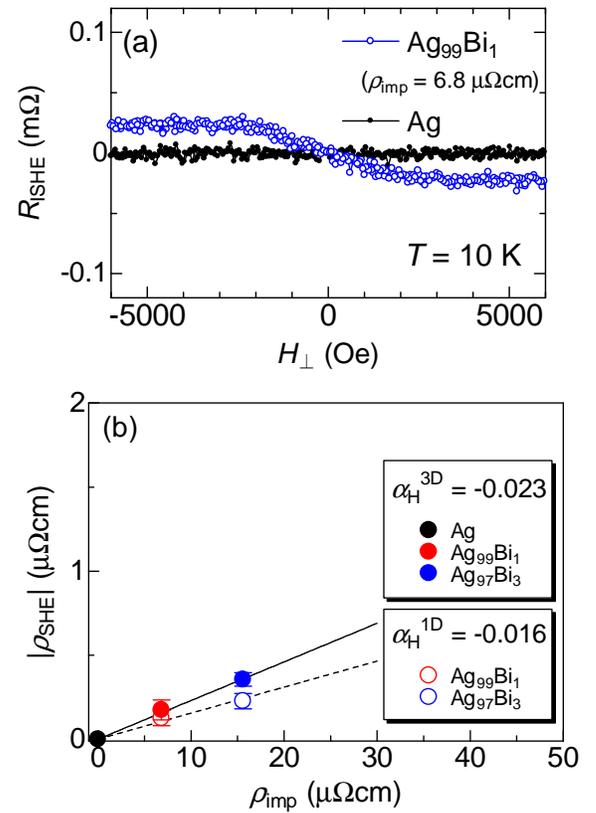}
\caption{(Color online) (a) $R_{\rm ISHE}$ of Ag$_{99}$Bi$_{1}$ measured at $T = 10$~K. As a reference signal, $R_{\rm ISHE}$ of pure Ag is also plotted. (b) SH resistivity $|\rho_{\rm SHE}|$ of AgBi alloys as a function of $\rho_{\rm imp}$. Closed and open symbols are data analyzed with the 3D and 1D models, respectively. The slopes of solid and broken lines correspond to the SH angles $\alpha_{\rm H}^{\rm 3D}$ and $\alpha_{\rm H}^{\rm 1D}$, respectively.} \label{fig9}
\end{center}
\end{figure}


\begin{figure}
\begin{center}
\includegraphics[width=8.5cm]{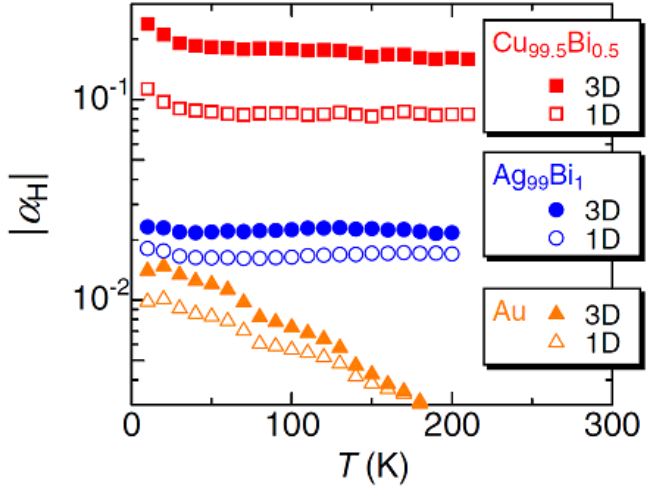}
\caption{(Color online) Temerature dependence of SH angles $|\alpha_{\rm H}|$ of Cu$_{99.5}$Bi$_{0.5}$ (square), Ag$_{99}$Bi$_{1}$ (circle), and Au (triangle) analyzed with the 3D (closed symbols) and 1D (open symbols) models.} \label{fig11}
\end{center}
\end{figure}

\begin{figure}
\begin{center}
\includegraphics[width=8cm]{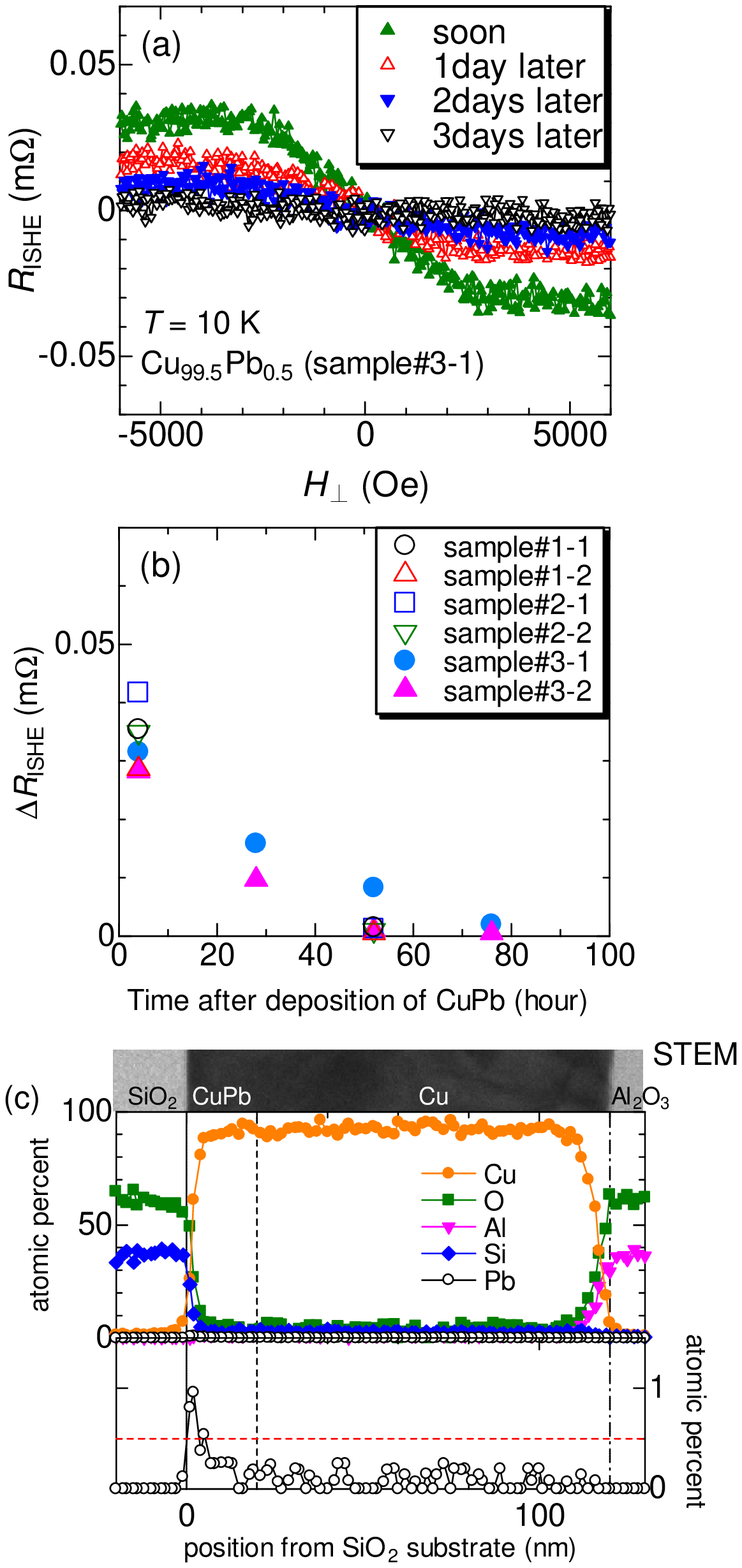}
\caption{(Color online) (a) Aging effect of $R_{\rm ISHE}$ of Cu$_{99.5}$Pb$_{0.5}$ measured at $T = 10$~K. (b) $\Delta R_{\rm ISHE}$ as a function of time for six different samples from three different batches. The open symbols are data measured right after the deposition of CuPb and two days later. The closed symbols are data measured everyday after the deposition of CuPb. (c) STEM image (upper panel) and EDX analyses of Cu$_{99.5}$Pb$_{0.5}$/Cu junction measured several days after the deposition of CuPb wire. The lower panel shows a closeup view for Pb. The vertical solid, broken, and dashed-dotted lines are the positions of interfaces between the substrate and CuPb, between CuPb and Cu, and surface of the Cu bridge, respectively. The transverse broken line in the lower panel is the Pb concentration of our CuPb target.} \label{fig12}
\end{center}
\end{figure}

In this subsection, we discuss the extrinsic SHEs measured with 
other material combinations. 
We first substituted Ag for the Cu bridge 
as illustrated in Fig.~\ref{fig8}(a). 
Figure~\ref{fig8}(b) shows $R_{\rm ISHE}$ 
of a 20~nm thick Cu$_{99.5}$Bi$_{0.5}$ wire 
bridged by a 100 nm thick Ag wire. 
As for the amplitude $\Delta R_{\rm ISHE}$, 
it is slightly smaller than that in Fig.~\ref{fig3}(a). 
This might be related to a slightly smaller spin diffusion length 
of a 100~nm thick Ag wire compared to Cu~\cite{idzuchi_apl_2012}, 
although we did not check the $L$ dependence of the NLSV signal without 
any middle wire. 
The sign of $R_{\rm ISHE}$ is also negative for $H(\theta=90^{\circ})$. 
Note that $H_{\perp}$ is defined as $H(\theta=90^{\circ})$.
We then changed the magnetic field angle to confirm if 
the observed signal for the Ag bridge 
really comes from the ISHE of the Cu$_{99.5}$Bi$_{0.5}$ wire.
The signal disappears when $\theta=0^{\circ}$ [see Fig~\ref{fig8}(b)]
and the sign reverses when $\theta=-90^{\circ}$. 
These results clearly show that $R_{\rm ISHE}$ does not 
depend on the bridge material which is 
used to transfer the pure spin current to the middle wire.

Next we keep the Cu bridge and change the host metal 
from Cu to Ag since Bi-doped Ag is also predicted to 
have a large SHE~\cite{gradhand_prl_2010,gradhand_prb_2010}. 
In Fig.~\ref{fig9}(a), we show $R_{\rm ISHE}$ of Ag$_{99}$Bi$_{1}$ 
and pure Ag measured at $T=10$~K. 
Apparently, when there is no Bi impurity in Ag, 
no ISHE is observed, which means that 
$R_{\rm ISHE}$ is negligibly small in pure Ag. 
This result is consistent with a recent spin pumping measurement with 
Ag/Py bilayer films~\cite{vila_nat_commun_2013}.
Once a small amount of Bi impurities is added in Ag, 
a negative ISHE is observed as shown in Fig.~\ref{fig9}(a). 
Contrary to the theoretical 
predictions~\cite{gradhand_prl_2010,gradhand_prb_2010}, however,
the observed ISHE of Ag$_{99}$Bi$_{1}$ was not so large as 
that of Cu$_{99.5}$Bi$_{0.5}$ [see Fig.~\ref{fig1}(a)]. 

To determine the SH angle of AgBi alloys, 
we plot $|\rho_{\rm SHE}|$ as a function of $\rho_{\rm imp}$. 
As can be seen in Fig.~\ref{fig9}(b), 
$|\rho_{\rm SHE}|$ linearly increases with increasing the 
Bi concentration up to 3\%, which clearly shows that 
the skew scattering is the dominant mechanism 
for the SHE~\cite{niimi_prl_2011}. 
This result also indicates that 
there is no segregation of Bi in Ag up to 3\%, while 
Bi impurities start to segregate from 0.5\% in Cu as 
already discussed in Ref.~\onlinecite{niimi_prl_2012}. 
Using the 1D and 3D models, $\alpha_{\rm H}^{1D}$ and 
$\alpha_{\rm H}^{\rm 3D}$ can be estimated as 
$-0.016$ and $-0.023$, respectively. 
Compared to CuBi alloys, the SH angle is one order smaller and 
comparable to that of CuIr alloys~\cite{niimi_prl_2011,niimi_prl_2012}.

So far, we have fixed the temperature at $T \leq 10$~K where 
the phonon contribution can be neglected. 
To see the temperature dependence of the SH angle is 
one of the best ways to discuss the dominant 
mechanism of the SHE~\cite{jin_prl_2009}. 
For this purpose, we measured $R_{\rm ISHE}$ and $R_{\rm S}$ every 10~K, 
and estimated $\alpha_{\rm H}^{\rm 1D}$ and $\alpha_{\rm H}^{\rm 3D}$. 
Figure~\ref{fig11} shows the temperature dependence of the absolute 
values of $|\alpha_{\rm H}^{\rm 1D}|$ and $|\alpha_{\rm H}^{\rm 3D}|$ 
for Cu$_{99.5}$Bi$_{0.5}$, Ag$_{99}$Bi$_{1}$, and Au. 
Ag$_{99}$Bi$_{1}$ has an almost temperature independent 
SH angle, which is characteristic of the skew 
scattering~\cite{niimi_prl_2011,jin_prl_2009}, and 
the difference between 
$|\alpha_{\rm H}^{\rm 1D}|$ and $|\alpha_{\rm H}^{\rm 3D}|$ 
is relatively small. 
Cu$_{99.5}$Bi$_{0.5}$ has 
almost the same tendency as Ag$_{99}$Bi$_{1}$ although 
there is a slight enhancement of the SH angle below 30~K.

As for Au, on the other hand, the temperature dependence of the 
SH angle is quite different from those of the other two alloys.
Both $\alpha_{\rm H}^{\rm 1D}$ and $\alpha_{\rm H}^{\rm 3D}$ 
decrease with increasing temperature and the difference between the two 
is getting smaller. 
The temperature dependence of the SHE of noble metal has been discussed 
in Pt~\cite{vila_prl_2007,kajiwara_j_phys_2009} as well as 
in Au~\cite{kajiwara_j_phys_2009}. 
In Ref.~\onlinecite{vila_prl_2007}, it was found that 
the SH conductivity of Pt is independent of temperature, 
which means that the SH angle increases with increasing temperature.
The same tendency has been confirmed in the inverse SH voltage for 
a Pt/Py bilayer film~\cite{kajiwara_j_phys_2009}. 
On the other hand, in the case of Au, 
the temperature dependence is opposite to 
the case of Pt~\cite{kajiwara_j_phys_2009}, 
which is consistent with the present result. 
In Refs.~\onlinecite{vila_prl_2007} and \onlinecite{kajiwara_j_phys_2009}, 
such a temperature dependence was attributed to the two extrinsic 
contributions, namely skew scattering and side jump. 
However, as already shown 
in a few theoretical~\cite{guo_prl_2008,tanaka_prb_2008} and 
experimental papers~\cite{jin_prl_2009}, 
the linearly increasing or decreasing SH angle with temperature is 
typical of the intrinsic mechanism based on the degeneracy of $d$ orbits 
by SO coupling. 
In the present case, the intrinsic mechanism is predominant since 
the SH angle decreases almost linearly with increasing temperature, which 
is qualitatively different from Ref.~\onlinecite{seki_nat_mater_2008} where 
the SH angle of Au is independent of temperature, 
characteristic of the skew scattering. 

We now come back to the Cu host and change the impurity. 
What happens when another 6$p$ impurity is doped in Cu? 
We fabricated SH devices using Cu$_{99.5}$Pb$_{0.5}$ and 
measured the ISHE and NLSV at $T=10$~K. 
In Fig.~\ref{fig12}(a), we show $R_{\rm ISHE}$ of Cu$_{99.5}$Pb$_{0.5}$. 
Note that all Cu$_{99.5}$Pb$_{0.5}$ samples were prepared 
within five hours after the deposition of CuPb middle wires. 
When we measure the Cu$_{99.5}$Pb$_{0.5}$ samples 
right after the deposition of the Cu bridge, 
it has two times smaller $\Delta R_{\rm ISHE}$ 
than Cu$_{99.5}$Bi$_{0.5}$, while the spin diffusion length of 
Cu$_{99.5}$Pb$_{0.5}$ is comparable or even larger than 
that of Cu$_{99.5}$Bi$_{0.5}$ [see Table~\ref{table1}]. 
From the ISHE and NLSV measurements, 
$\alpha_{\rm H}^{\rm 3D}$ of Cu$_{99.5}$Pb$_{0.5}$ is estimated to be 
$-0.13$, which is half of $\alpha_{\rm H}^{\rm 3D}$ of Cu$_{99.5}$Bi$_{0.5}$. 

What is interesting to note is the aging effect of $R_{\rm ISHE}$ of 
Cu$_{99.5}$Pb$_{0.5}$. After the measurements at $T=10$~K, 
the samples were warmed up to room temperature 
and kept in a vacuum box for 1 day. 
Then we measured the same samples again at $T=10$~K. 
As can be seen in Fig.~\ref{fig12}(a), $R_{\rm ISHE}$ of 
Cu$_{99.5}$Pb$_{0.5}$ decreases by more than half. 
We simply repeated this procedure until we could not obtain $R_{\rm ISHE}$.
In a few days, $R_{\rm ISHE}$ of Cu$_{99.5}$Pb$_{0.5}$ disappeared.
On the other hand, Cu$_{99.5}$Bi$_{0.5}$ 
did not show such an aging effect at least in one week. 
In Fig.~\ref{fig12}(b), we plot $\Delta R_{\rm ISHE}$ of Cu$_{99.5}$Pb$_{0.5}$ 
as a function of time after the deposition of the CuPb wire.
For six samples from three different batches, 
$\Delta R_{\rm ISHE}$ disappears within a few days. 

The above result indicates that Pb impurities 
in the middle wire move somewhere. 
In fact, it is known that Pb has a very fast mobility. 
According to a scanning tunneling microscopy study on 
Cu islands on Pb(111) substrate~\cite{stm_lead}, Cu islands are masked by 
Pb atoms from the substrate. 
This migration originates from a very fast mobility of Pb in Cu.
We have taken a scanning tunneling electron microscopy (STEM)
image of the Cu$_{99.5}$Pb$_{0.5}$ and Cu junction and performed 
energy dispersive x-ray (EDX) analyses several days after 
the deposition of CuPb wire. 
In our procedure, we cannot take the STEM image right after 
the CuPb deposition, but as shown in Fig.~\ref{fig12}(c), 
Pb atoms are segregated at the bottom of the middle wire several days after 
the fabrication. 
Presumably, such segregated Pb impurities at the substrate 
would not contribute to the ISHE, and thus $R_{\rm ISHE}$ of 
Cu$_{99.5}$Pb$_{0.5}$ becomes almost zero in a few days.

Before closing this subsection, we discuss the reason why 6$p$ impurities 
such as Pb and Bi in the Cu host show large SH angles. 
First of all, Pb and Bi have large SO interactions 
because of their large atomic numbers. 
According to recent theoretical 
predictions~\cite{gradhand_prl_2010,gradhand_prb_2010}, 
a large difference of SO interactions between 
the host (in the present case, Cu) and impurity metals is 
one of the key points to have a large SHE based on 
the skew scattering mechanism. 
In addition, not only the atomic number but also the orbit of outermost shell 
play an important role in the strength of SO interaction~\cite{orbit_vs_SO}. 
Intuitively thinking, when the orbital angular momentum is large, 
the SO interaction is also large. 
However, to obtain the expectation value of SO interaction, 
one needs to multiply the wave functions. 
When the orbital angular momentum is large ($d$ or $f$ orbits), 
the wave function is rather closed and does not have a chance 
to hybridize with the most outer orbit. 
On the other hand, $p$ orbits have spread wave functions and thus have 
larger SO interactions~\cite{orbit_vs_SO}. 
Therefore, 6$p$ impurities such as Pb and Bi 
have the largest SO interactions. 
For CuBi, a calculation based on a phase shift model predicts 
a negative SH angle but smaller ($-0.046$) than the experimental one. 
A similar calculation has not been performed for AgBi and CuPb. 

\subsection{SHE of Ta}

Finally, let us mention the spin diffusion length and the SH angle of Ta. 
This topic was the cause of the big debate 
among several groups~\cite{liu_arxiv_2011}.
In Ref.~\onlinecite{morota_prb_2011}, some of 
the present authors reported 
$\lambda_{\rm M}^{\rm 1D}$ and $\alpha_{\rm H}^{\rm 1D}$ of Ta
using the 1D models, i.e., 
Eqs.~(\ref{eq2}) and (\ref{eq3}), respectively. 
However, according to Ref.~\onlinecite{liu_science_2012}, 
the SH angle of Ta is $-0.15$, which 
is about 40 times larger than our $\alpha_{\rm H}^{\rm 1D}$ of Ta. 
Liu~\textit{et al}. pointed out in Ref.~\onlinecite{liu_arxiv_2011} that 
the overestimations of the shunting factor $x$ 
and $\lambda_{\rm M}^{\rm 1D}$ result in a large underestimation 
of the SH angle of Ta. 
Thus, we have performed the 3D analysis to obtain 
$\lambda_{\rm M}^{\rm 3D}$ and $\alpha_{\rm H}^{\rm 3D}$ of Ta. 
As can be seen in Table~\ref{table1}, $\lambda_{\rm M}^{\rm 3D}$ is 
3~nm, which is the same as $\lambda_{\rm M}^{\rm 1D}$. 
On the other hand, $\alpha_{\rm H}^{\rm 3D}$ is $-0.008$ which is 
twice larger than $\alpha_{\rm H}^{\rm 1D}$. 
This difference certainly comes from the overestimation of $x$.
However, the large SH angle reported in Ref.~\onlinecite{liu_science_2012} 
cannot be reproduced in our analysis. 
The big difference in $\alpha_{\rm H}$ of Ta 
between Ref.~\onlinecite{liu_science_2012} and the present result is 
not clear yet, but apparently it does not originate from 
the overestimation of the spin diffusion length of Ta. 
We believe that in the ferromagnetic/Ta bilayer system, 
some additional effects are induced at the interface between Ta and 
the ferromagnet~\cite{ohno_nat_mater_2013}, 
and thus the SH angle obtained in Ref.~\onlinecite{liu_science_2012} 
is seemingly enhanced by a factor of more than 10.

\section{Conclusions}

We have experimentally studied the extrinsic SHEs of 
CuBi, AgBi, and CuPb alloys 
with the spin absorption technique in the lateral spin valve structure. 
Among them, CuBi shows the largest SHE. 
The SH angle estimated with 
the 3D model amounts to $-0.24$. 
Such a large SHE of CuBi has been supported 
by several additional measurements; 
(i) comparison with the SHE of Au which has a 
comparable spin diffusion length to CuBi, 
(ii) WAL measurements to obtain the spin diffusion length, 
(iii) the CuBi thickness dependence of the SH angle, 
(iv) ISHE measurements with substitution of Ag for the Cu bridge, and 
(v) the magnetic field angle dependence of the ISHEs. 
CuPb also shows a large SH angle ($-0.13$) but the SH signal disappears 
in a few days presumably because of the fast migration of Pb in Cu. 
On the other hand, when the Cu host is replaced with Ag, 
the SH angle is reduced by a factor of ten. 

\begin{acknowledgments}
We acknowledge helpful discussions with S.~Maekawa, B.~Gu, J.~Bass, 
W.~P.~Pratt, T.~Kato, Y.~Yanase, H.~Harima, and K.~Kondou. 
We would also like to thank Y.~Iye and S.~Katsumoto 
for the use of the lithography facilities. 
This work was supported by KAKENHI 
(Grant No. 22840012, 24740217, and 23244071).
\end{acknowledgments}

\end{document}